\documentclass[12pt,compress,nosort]{article}
\usepackage{amsmath,axodraw,epsf,graphicx,booktabs,subfigure}

\def\Year{\expandafter\eatPrefix\the\year}
\newcount\hours \newcount\minutes
\def\monthname{\ifcase\month\or
January\or February\or March\or April\or May\or June\or July\or
August\or September\or October\or November\or December\fi}
\def\shortmonthname{\ifcase\month\orx
Jan\or Feb\or Mar\or Apr\or May\or Jun\or Jul\or
Aug\or Sep\or Oct\or Nov\or Dec\fi}

\def\TimeStamp{\hours\the\time\divide\hours by60%
\minutes -\the\time\divide\minutes by60\multiply\minutes by60%
\advance\minutes by\the\time%
${\rm \shortmonthname}\cdot   \if\day<10{}0\fi\the\day\cdot   \the\year
\qquad\the\hours:\if\minutes<10{}0\fi\the\minutes$}

\newskip\humongous \humongous=0pt plus 1000pt minus 100pt
\def\caja{\mathsurround=0pt}
\def\eqalign#1{\,\vcenter{\openup1\jot \caja
       \ialign{\strut \hfil$\displaystyle{##}$&$
        \displaystyle{{}##}$\hfil\crcr#1\crcr}}\,}
\newif\ifdtup

\newcounter{eqnumber}[section]
\renewcommand{\theeqnumber}{\thesection.\arabic{eqnumber}}
\def\equn{\refstepcounter{eqnumber}
\eqno({\rm \theeqnumber})
}

\def\npb#1#2#3{{\rm Nucl. Phys. B}{\bf \ #1}, #3 (#2)}

\def\hepph#1{[hep-ph/#1]}

\def\li{{\rm Li}_2}

\def\Fs#1#2{F^{{#1}}_{#2}}
\def\Fone{\Fs{\rm 1m}}
\def\Feasy{\Fs{{\rm 2m}\,e}}
\def\Fhard{\Fs{{\rm 2m}\,h}}

\def\Wsix#1{W_6^{(#1)}}

\newbox\charbox
\newbox\slabox
\def\s#1{{      
        \setbox\charbox=\hbox{$#1$}
        \setbox\slabox=\hbox{$/$}
        \dimen\charbox=\ht\slabox
        \advance\dimen\charbox by -\dp\slabox
        \advance\dimen\charbox by -\ht\charbox
        \advance\dimen\charbox by \dp\charbox
        \divide\dimen\charbox by 2
        \raise-\dimen\charbox\hbox to \wd\charbox{\hss/\hss}
        \llap{$#1$}
}}

\def\spa#1.#2{\left\langle#1\,#2\right\rangle}
\def\spb#1.#2{\left[#1\,#2\right]}
\def\lor#1.#2{\left(#1\,#2\right)}

\catcode`@=11  

\def\eps{\epsilon}

\def\e{\epsilon}

\def\ra{\rangle}

\def\lsl{\not{\hbox{\kern-2.3pt $\ell$}}}
\def\ksl{\not{\hbox{\kern-2.3pt $k$}}}

\def\rg{r_{\Gamma}}

\def\spa#1.#2{\left\langle#1\,#2\right\rangle}
\def\spb#1.#2{\left[#1\,#2\right]}
\def\lor#1.#2{\left(#1\,#2\right)}

\def\sand#1.#2.#3{%
  \left\langle\smash{#1}{\vphantom1}\right|{#2}%
  \left|\smash{#3}{\vphantom1}\right\rangle}
\def\sandp#1.#2.#3{%
  \left\langle\smash{#1}{\vphantom1}^{-}\right|{#2}%
  \left|\smash{#3}{\vphantom1}^{+}\right\rangle}
\def\sandpp#1.#2.#3{%
  \left\langle\smash{#1}{\vphantom1}^{+}\right|{#2}%
  \left|\smash{#3}{\vphantom1}^{+}\right\rangle}
\def\sandmm#1.#2.#3{%
  \left\langle\smash{#1}{\vphantom1}^{-}\right|{#2}%
  \left|\smash{#3}{\vphantom1}^{-}\right\rangle}
\def\sandpm#1.#2.#3{%
  \left\langle\smash{#1}{\vphantom1}^{+}\right|{#2}%
  \left|\smash{#3}{\vphantom1}^{-}\right\rangle}
\def\sandmp#1.#2.#3{%
  \left\langle\smash{#1}{\vphantom1}^{-}\right|{#2}%
  \left|\smash{#3}{\vphantom1}^{+}\right\rangle}

\def\Atree{A^{\rm tree}}

\def\LB{\left[}\def\RB{\right]}

\def\tn#1#2{t^{[#1]}_{#2}}

\def\tree{{\rm tree}}

\def\NeqEight{{\cal N} = 8}

\def\NeqFour{{\cal N} = 4}
\def\NeqOne{{\cal N} = 1}

\newcommand\mmpppp{\text{$-$$-$$+$$+$$+$$+$}}
\newcommand\mmmppp{\text{$-$$-$$-$$+$$+$$+$}}
\newcommand\mmpmpp{\text{$-$$-$$+$$-$$+$$+$}}
\newcommand\mpmpmp{\text{$-$$+$$-$$+$$-$$+$}}

\newcommand\mmppppp{\text{$-$$-$$+$$+$$+$$+$$+$}}
\newcommand\mmmpppp{\text{$-$$-$$-$$+$$+$$+$$+$}}
\newcommand\mmpmppp{\text{$-$$-$$+$$-$$+$$+$$+$}}
\newcommand\mmppmpp{\text{$-$$-$$+$$+$$-$$+$$+$}}
\newcommand\mpmpmpp{\text{$-$$+$$-$$+$$-$$+$$+$}}

\newcommand\mmpppppp{\text{$-$$-$$+$$+$$+$$+$$+$$+$}}
\newcommand\mmmppppp{\text{$-$$-$$-$$+$$+$$+$$+$$+$}}
\newcommand\mmpmpppp{\text{$-$$-$$+$$-$$+$$+$$+$$+$}}
\newcommand\mmppmppp{\text{$-$$-$$+$$+$$-$$+$$+$$+$}}
\newcommand\mpmpmppp{\text{$-$$+$$-$$+$$-$$+$$+$$+$}}
\newcommand\mpmppmpp{\text{$-$$+$$-$$+$$+$$-$$+$$+$}}

\begin{document}

\begin{titlepage}

\begin{flushright}
\today
\\
hep-th/yymmnnn
\\

\end{flushright}

\vskip 2.cm

\begin{center}
\begin{Large}
{\bf Approximate Universality of $\NeqFour$ Super-Yang--Mills One-Loop Amplitudes}

\vskip 2.cm

\end{Large}

\vskip 2.cm
{\large
David~C.~Dunbar${}$, James H. Ettle and
Warren~B.~Perkins
} 

\vskip 0.5cm

\vskip 0.5cm

{\it  Department of Physics \\
Swansea University\\
 Swansea, SA2 8PP, UK }

\vskip .3cm

\begin{abstract}

A study of the colour-ordered one-loop amplitudes in $\NeqFour$ SYM reveals a surprising property: numerically, the amplitudes have an Approximately Universal form
$A^{\text{1-loop}}_n \sim A^{\text{tree}}_n \times U_n$, where $U_n$ is helicity independent.  This form is exact if the amplitude is a MHV amplitude,  but has an
``error'' which is typically less than 1\% for the six, seven and eight gluon NMHV one-loop amplitudes.  

\end{abstract}

\end{center}

\vfill

\end{titlepage}

\section{Approximate Universality}

The maximally supersymmetric Yang--Mills theory, $\NeqFour$, is an extremely special four dimensional field theory. 
In this letter we demonstrate that the colour ordered one-loop scattering amplitudes are remarkably close to a Universal form:
$$
A^{\text{1-loop}}_n = A^{\rm tree}_n  U_n  +{\cal E}_{H,n} \, ,
\equn
$$
where $U_n$ is helicity-independent and ${\cal E}_{H,n}$ is a helicity-dependent correction. $U_n$ contains the exact $\eps^{-2}$ and $\eps^{-1}$ infrared singular terms. 
We find that the error is surprisingly small, being less than $1\%$ for the 
explicit six, seven and eight point amplitudes we study.  

The exact analytic forms of many one-loop amplitudes are known, however Approximate Universality is not manifest from an analytic study of these expressions. 
Instead it
is observed when we numerically evaluate these analytic forms.  In the next section we review the known analytic forms of $\NeqFour$ one-loop $n$-point amplitudes and in  section 3 we present a numerical study of these expressions for six and seven gluon amplitudes together with the NMHV eight-point amplitudes.

\section{General form of $\NeqFour$ one-loop amplitudes}

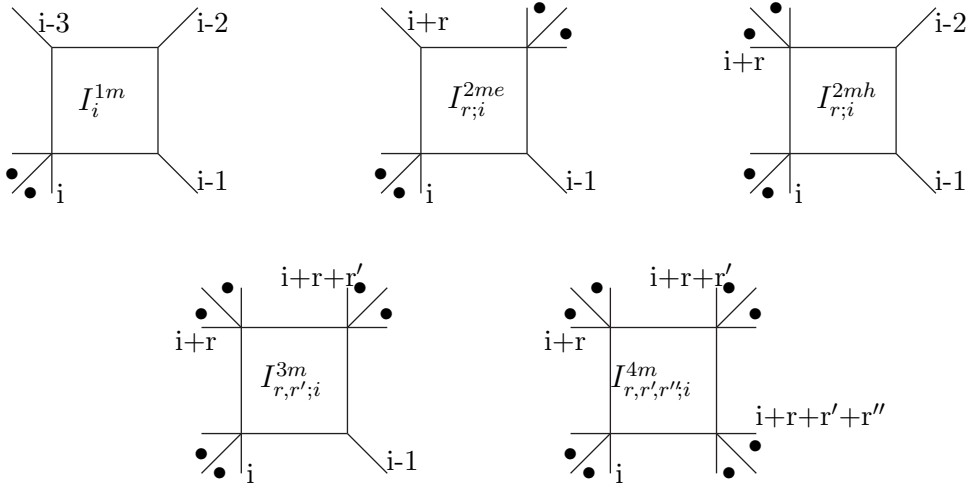
\begin{figure}[h]
\centering\subfigure{
\begin{picture}(120,95)(0,0)
\Line(30,20)(70,20)
\Line(30,60)(70,60)
\Line(30,20)(30,60)
\Line(70,60)(70,20)                                     
\Line(30,20)(15,5)
\Line(30,60)(15,75)
\Line(70,20)(85,5)
\Line(70,60)(85,75)                                                         
\Line(30,20)(30,5)
\Line(30,20)(15,20)
\Text(22,5)[c]{\small$\bullet $}
\Text(15,12)[c]{\small $\bullet $}                                    
\Text(32,5)[l]{\small $\hbox{\rm i}$}
\Text(85,10)[l]{\small $\hbox{\rm i-1}$}
\Text(85,70)[l]{\small $\hbox{\rm i-2}$}
\Text(25,70)[l]{\small $\hbox{\rm i-3}$}                                   
\Text(40,40)[l]{$I^{1m}_{i}$}
\end{picture}
}\quad\subfigure{
\begin{picture}(120,95)(0,0)
\Line(30,20)(70,20)
\Line(30,60)(70,60)
\Line(30,20)(30,60)
\Line(70,60)(70,20)                                                                   
\Line(30,20)(15,5)
\Line(30,60)(15,75)
\Line(70,20)(85,5)
\Line(70,60)(85,75)                                                             
\Line(30,20)(30,5)
\Line(30,20)(15,20)
\Text(22,5)[c]{\small$\bullet$}
\Text(15,12)[c]{\small $\bullet$}                                                             
\Line(70,60)(70,75)
\Line(70,60)(85,60)
\Text(75,75)[c]{\small$\bullet$}
\Text(85,65)[c]{\small $\bullet$}                                               
\Text(32,5)[l]{\small $\hbox{\rm i}$}
\Text(85,10)[l]{\small $\hbox{\rm i-1}$}
\Text(25,70)[l]{\small $\hbox{\rm i+r}$}                                
\Text(40,40)[l]{$I^{2me}_{r;i}$}
\end{picture}
}\quad\subfigure{
\begin{picture}(120,95)(0,0)
\Line(30,20)(70,20)
\Line(30,60)(70,60)
\Line(30,20)(30,60)
\Line(70,60)(70,20)                                  
\Line(30,20)(15,5)
\Line(30,60)(15,75)
\Line(70,20)(85,5)
\Line(70,60)(85,75)                                           
\Line(30,20)(30,5)
\Line(30,20)(15,20)
\Text(22,5)[c]{\small$\bullet$}
\Text(15,12)[c]{\small $\bullet$}                                           
\Line(30,60)(30,75)
\Line(30,60)(15,60)
\Text(15,65)[c]{\small$\bullet$}
\Text(25,75)[c]{\small $\bullet$}                                        
\Text(32,5)[l]{\small $\hbox{\rm i}$}
\Text(85,10)[l]{\small $\hbox{\rm i-1}$}
\Text(85,70)[l]{\small $\hbox{\rm i-2}$}
\Text(5,55)[l]{\small $\hbox{\rm i+r}$}                                      
\Text(40,40)[l]{$I^{2mh}_{r;i}$}
\end{picture}
}\\\subfigure{
\begin{picture}(120,95)(0,0)
\Line(30,20)(70,20)
\Line(30,60)(70,60)
\Line(30,20)(30,60)
\Line(70,60)(70,20)                                           
\Line(30,20)(15,5)
\Line(30,60)(15,75)
\Line(70,20)(85,5)
\Line(70,60)(85,75)                                                        
\Line(30,20)(30,5)
\Line(30,20)(15,20)
\Text(22,5)[c]{\small $\bullet$}
\Text(15,12)[c]{\small $\bullet$}                                            
\Line(30,60)(30,75)
\Line(30,60)(15,60)
\Text(15,65)[c]{\small $\bullet$}
\Text(25,75)[c]{\small $\bullet$}                                         
\Line(70,60)(70,75)
\Line(70,60)(85,60)
\Text(75,75)[c]{\small $\bullet$}
\Text(85,65)[c]{\small $\bullet$}                                                                                                                           
\Text(32,5)[l]{\small $\hbox{\rm i}$}
\Text(85,10)[l]{\small $\hbox{\rm i-1}$}
\Text(45,80)[l]{\small $\hbox{\rm i+r+r}'$}
\Text(5,55)[l]{\small $\hbox{\rm i+r}$}                                    
\Text(37,40)[l]{$I^{3m}_{r,r';i}$}
\end{picture}
}\quad\subfigure{
\begin{picture}(120,95)(0,0)
\Line(30,20)(70,20)
\Line(30,60)(70,60)
\Line(30,20)(30,60)
\Line(70,60)(70,20)                                     
\Line(30,20)(15,5)
\Line(30,60)(15,75)
\Line(70,20)(85,5)
\Line(70,60)(85,75)                                            
\Line(30,20)(30,5)
\Line(30,20)(15,20)
\Text(22,5)[c]{\small $\bullet$}
\Text(15,12)[c]{\small $\bullet$}                                            
\Line(30,60)(30,75)
\Line(30,60)(15,60)
\Text(15,65)[c]{\small $\bullet$}
\Text(25,75)[c]{\small $\bullet$}                                                                                                                          
\Line(70,60)(70,75)
\Line(70,60)(85,60)
\Text(75,75)[c]{\small $\bullet$}
\Text(85,65)[c]{\small $\bullet$}                                             
\Line(70,20)(70,5)
\Line(70,20)(85,20)
\Text(75,5)[c]{\small $\bullet$}
\Text(85,15)[c]{\small $\bullet$}                                         
\Text(32,5)[l]{\small $\hbox{\rm i}$}
\Text(85,27)[l]{\small $\hbox{\rm i+r+r}'\hbox{\rm +r}''$}
\Text(45,80)[l]{\small $\hbox{\rm i+r+r}'$}
\Text(5,55)[l]{\small $\hbox{\rm i+r}$}                                     
\Text(30,40)[l]{$I^{4m}_{r,r'\hskip -2pt ,r''\hskip -2pt;i}$}
\end{picture}}
\caption{The scalar box functions with their labeling.}
\label{BoxesFigure}
\end{figure}

The one-loop amplitudes for gluon scattering in $\NeqFour$ super-Yang--Mills are particularly simple, being heavily constrained by the large symmetry. They can be expressed entirely in terms of scalar boxes~\cite{BDDKa}:
$$
 A_n^{\NeqFour}=\sum_{a\in \cal C}\, c_a\, I^{a}_4   \ .
\equn\label{boxexpansion} 
$$
The amplitude is completely defined by specifying the rational coefficients $c_a$. This is a considerable simplification relative to a generic Yang--Mills amplitude which would contain scalar triangle functions, scalar bubble functions and additional rational pieces.
The set of scalar box functions is
 $$
  I_{i}^{1{\rm m}}
\hskip 0.5truecm
 I_{r;i}^{2{\rm m}e}
\hskip 0.5truecm
  I_{r;i}^{2{\rm m}h}
\hskip 0.5truecm
  I_{r,r';i}^{3{\rm m}}
\hskip 0.5truecm
I_{r, r', r''; i}^{4{\rm m}}, \equn
$$
with the labeling as indicated in figure~\ref{BoxesFigure}.

 Such 
 amplitudes are  ``cut-constructible''  meaning  they may be reconstructed using cutting rules
with on-shell four dimensional tree amplitudes~\cite{BDDKa}.  
The coefficients $c_a$ may be determined in various ways,
originally via the two-particle cuts~\cite{BDDKa,BDDKb}, but perhaps most conveniently using ``quadruple cuts''~\cite{BrittoUnitarity}. 
The quadruple cuts give the 
box-coefficients as a product of four tree amplitudes,
$$
\eqalign{
c_{\text{box}}
={ 1 \over 2 } \sum_{\cal S}
\biggl( \Atree(\ell_1^{s_1},i_1,  & \ldots,i_2,-\ell_2^{-s_2}) \times
\Atree(\ell_2^{s_2},i_3,\ldots,i_4,-\ell_3^{-s_3})\times 
\cr
& \hspace{0.1 cm}
\Atree(\ell_3^{s_3},i_5,\ldots,i_6,-\ell_4^{-s_4}) \times
\Atree(\ell_4^{s_4},i_7,\ldots,i_8,-\ell_1^{-s_1}) \biggr)\;.
\cr}
\equn\label{QuadCuts}
$$ 
The sum is over all allowed intermediate configurations and
particle types~\cite{BrittoUnitarity}. Using the expansion of eq.~(\ref{boxexpansion}) together with the quadruple cuts makes the computation of one-loop amplitudes
considerably more tractable than the corresponding QCD amplitudes.  

It is convenient to define rescaled box functions $F^{a}$ by
$$
\eqalign{
  I_{i}^{1{\rm m}} 
=
\ -2 \rg {\Fone{i} \over \tn{2}{i-3} \tn{2}{i-2} }  \;\;\;
 I_{r;i}^{2{\rm m}e}
&=
\ -2 \rg {\Feasy{r;i}
      \over \tn{r+1}{i-1}\tn{r+1}{i} -\tn{r}{i}\tn{n-r-2}{i+r+1} }\,, \;\;\;
I_{r;i}^{2{\rm m}h}
=\ -2 \rg {\Fhard{r;i} \over \tn{2}{i-2} \tn{r+1}{i-1} } \,,
\cr
 I_{r,r';i}^{3{\rm m}}
&=\ -2 \rg {F^{3\rm m}_{r,r';i}
     \over \tn{r+1}{i-1} \tn{r+r'}i -\tn{r}{i} \tn{n-r-r'-1}{i+r+r'} }\,.
 \cr}
\equn
$$
The $F^a$ functions  are the box functions with the appropriate momentum prefactors removed; explicit formulae may be found in, for example, ref.\cite{BDDKa}.  
The momentum invariants are 
$$\eqalign{
\tn{r}{i} &= (k_i+\cdots +k_{i+r-1})^2,
}
\equn$$
and $\rg=\Gamma(1+\eps)\Gamma^2(1-\eps)/\Gamma(1-2\eps)$. Analytic results will be given in the
four-dimensional helicity regularisation scheme (FDH)~\cite{Bern:1991aq}
with the box integrals defined in $4-2\eps$ dimensions.  The amplitudes in other schemes such as 't~Hooft--Veltman are simply related to those in FDH via
$$
A^{\text{1-loop}}_{\text{FDH}} = 
A^{\text{1-loop}}_{\text{'tHV}} + { A^{\tree} \over 3(4\pi)^2  }  ,
\equn
$$
which leads to a small modification of $U_n$ but leaves ${\cal E}_{H,n}$ unchanged.

In any supersymmetric theory the gluon scattering amplitudes with all or all-but-one helicities of the same type vanish
$$
A_n^{\NeqFour}(1^+,2^+,\dots ,n^+)
=
A_n^{\NeqFour}(1^-,2^+,\dots ,n^+)
=0,
\equn
$$
and the simplest non-vanishing amplitudes are the ``MHV'' (Maximally Helicity-Violating) amplitudes which have exactly two negative helicity gluons. 
At tree level these amplitudes are given by the Parke--Taylor formula~\cite{ParkeTaylor}, 
$$
\eqalign{
  \Atree_n(1^+,\ldots,j^-,\ldots,k^-,
                \ldots,n^+),
&=\ i\, { {\spa{j}.{k}}^4 \over \spa1.2\spa2.3\cdots\spa{n}.1 }\ ,
  \cr}
\equn\label{ParkeTaylor}
$$ for a partial amplitude where $j$ and $k$ label the legs with
negative helicity.  We use the notation  
$\spa{j}.{l}\equiv \langle j^- | l^+ \rangle $, 
$\spb{j}.{l} \equiv \langle j^+ |l^- \rangle $, 
with $| i^{\pm}\ra $ 
being a massless Weyl spinor with
momentum $k_i$ and chirality
$\pm$~\cite{SpinorHelicity,ManganoReview}.  The spinor products are
related to momentum invariants by 
$\spa{i}.j\spb{j}.i=2k_i \cdot k_j\equiv s_{ij}$ 
with $\spa{i}.j^*=\spb{j}.i$.

The one-loop MHV amplitudes in $\NeqFour$ have the form~\cite{BDDKa}:
\def\cg{{ r_{\Gamma} \over (4\pi)^{2-\eps}}}
$$
A_n^{\NeqFour,MHV}=
\cg \Atree_n \times \biggl(  \sum_i F_i^{1m}   
+\sum_{i,r}  F_{r;i}^{2me}  
\biggr)
= \cg \Atree_n \times  U_n,
\equn
$$
where the sum of $F$ functions is over all possible inequivalent 
functions with the appropriate cyclic ordering of legs. 
The form $U_n$ is universal in that it does not depend upon which two legs have negative helicity.  
The sum over box coefficients can be expressed in terms of logarithms and dilogarithms via
$$
U_n 
=
\sum_{i=1}^{n}  -{ 1 \over \eps^2 } \biggl(
{ \mu^2  \over -\tn2{i} } \biggr)^{\eps}
-\sum_{r=2}^{\lfloor n/2\rfloor -1}
\sum_{i=1}^n
  \ln \biggl({ -\tn{r}{i}\over -\tn{r+1}{i} }\biggr)
  \ln \biggl({ -\tn{r}{i+1}\over -\tn{r+1}{i} }\biggr) +
D_n + L_n +{ n \pi^2 \over 6 }\ ,
\equn\label{UniversalFunc}
$$
where all indices are understood to be mod $n$ and $\mu^2$ is the renormalisation scale. 
The forms of $D_n$ and $L_n$ depend upon whether $n$ is odd or even.
For $n=2m+1$,
$$
D_{2m+1}= -\sum_{r=2}^{m-1} \Biggl( \sum_{i=1}^{n}
\li \biggl[ 1- { \tn{r}{i} \tn{r+2}{i-1}
\over \tn{r+1}{i} \tn{r+1}{i-1} } \biggr]  \Biggr)\ ,
\equn
$$
$$
L_{2m+1}= -{ 1\over 2} \sum_{i=1}^n
  \ln \biggl({ -\tn{m}{i}\over -\tn{m}{i+m+1}  } \biggr)
  \ln \biggl({ -\tn{m}{i+1}\over -\tn{m}{i+m} } \biggr)\ ,
\equn
$$
whereas for $n=2m$,
$$
D_{2m}= -\sum_{r=2}^{m-2} \Biggl( \sum_{i=1}^{n}
\li \biggl[ 1- { \tn{r}{i} \tn{r+2}{i-1}
\over \tn{r+1}{i} \tn{r+1}{i-1} }  \biggr]  \Biggr)
-\sum_{i=1}^{n/2} \li \biggl[ 1- { \tn{m-1}{i} \tn{m+1}{i-1}
\over \tn{m}{i} \tn{m}{i-1}} \biggr]\ ,
\equn
$$
$$
L_{2m}=-{1\over 4} \sum_{i=1}^n
  \ln \biggl({ -\tn{m}{i}\over -\tn{m}{i+m+1}  } \biggr)
  \ln \biggl({ -\tn{m}{i+1}\over -\tn{m}{i+m} } \biggr)\ .
\hskip 4 truecm
\equn
$$

The NMHV amplitudes have also been calculated~\cite{BDDKb,Britto:2004nj,Bern:2004ky,BDKn} and have very different forms from the 
MHV amplitudes. For example let us examine the six-point amplitudes. 
The three independent NMHV amplitudes are constructed from  the one-mass box functions, $F_i^{1m}$ and the
two-mass-hard  boxes $F_i^{2m h}$ in the very special combinations
$$
\eqalign{
   \Wsix{i}\ &\equiv\  \Fone{i} + \Fone{i+3}
                 + \Fhard{2;i+1} + \Fhard{2;i+4}  \cr             
     &= -{1\over2\e^2} \sum_{j=1}^6
         \left( { \mu^2 \over -t_j^{[2]} } \right)^\e
   - \ln\left({-t_{i}^{[3]}\over -t_i^{[2]}}\right)
      \ln\left({-t_{i}^{[3]} \over -t_{i+1}^{[2]}}\right)
  \cr &\quad
   - \ln\left({-t_{i}^{[3]} \over -t_{i+3}^{[2]}}\right)
      \ln\left({-t_{i}^{[3]} \over -t_{i+4}^{[2]}}\right)
 + \ln\left({-t_{i}^{[3]} \over -t_{i+2}^{[2]}}\right)
      \ln\left({-t_{i}^{[3]} \over -t_{i+5}^{[2]}}\right)
      \cr &\quad
  + {1\over 2}\ln\left({-t_{i}^{[2]} \over -t_{i+3}^{[2]}}\right)
         \ln\left({-t_{i+1}^{[2]} \over -t^{[2]}_{i+4}}\right)
 + {1\over 2}\ln\left({-t_{i-1}^{[2]} \over -t_{i}^{[2]}}\right)
         \ln\left({-t_{i+1}^{[2]} \over -t_{i+2}^{[2]}}\right)
         \cr &\quad
 + {1\over 2}\ln\left({-t_{i+2}^{[2]}\over -t_{i+3}^{[2]}}\right)
         \ln\left({-t_{i+4}^{[2]} \over -t_{i+5}^{[2]} }\right)
  + {\pi^2\over3}\ . \cr}
\equn\label{Wdef}
$$
As we can see the dilogarithms drop out of this expression.  This feature of the $\NeqFour$ NMHV six-point 
amplitudes persists 
for amplitudes involving external states other than gluons~\cite{Bidder:2005in} but not beyond six points. 
The first NMHV $\NeqFour$ amplitude is given by
$$
\eqalign{
A_{6}^{\NeqFour}(& 1^-,2^-,3^-,4^+,5^+,6^+)\ 
=i\cg\ \LB B_1\,\Wsix1+B_2\,\Wsix2+B_3\,\Wsix3\RB
\cr},
\equn\label{pppmmmloop}
$$
where the coefficients $B_i$ are given in terms of the $B_0$ function
$$
\eqalign{
B_0\ &=\  {
  \,  t_{123}^3
  \over \spb1.2\spb2.3\spa4.5\spa5.6\
    [1|K_{23}|4\ra [3|K_{12}|6\ra }
  \cr}
\equn
$$
with
$$
\eqalign{
  B_1\ &=\ B_0 \ , \cr
  B_2\ &=\ \left({ [4|K_{123}|1\ra 
         \over t_{234} } \right)^4  \ B_0^+
       + \left({ \spa2.3\spb5.6 \over t_{234} } \right)^4
            \ \bar{B}_0^+
          \ , \cr
  B_3\ &=\ \left({ [6|K_{345}|3\ra 
         \over t_{345} } \right)^4  \ B_0^- 
       + \left({ \spa1.2\spb4.5 \over t_{345} } \right)^4
            \ \bar{B}_0^-   \cr}
\equn
$$
and
$$
B_0^+ \equiv B_0|_{i \longrightarrow i+1} \;\;
B_0^- \equiv B_0|_{i \longrightarrow i-1} 
\;\;\;\;\;
\bar{B}_0
\equiv
B_0 | _{ \spa{a}.b \leftrightarrow \spb{a}.b }.
\equn
$$
For convenience we use $t_{123}\equiv (k_1+k_2+k_3)^2$ and $[1|K_{23}|4\rangle = [1\:2]\langle 2\:4 \rangle + [1\:3]\langle3\:4\rangle$, etc. 

The other amplitudes are:
$$
\eqalign{
A_{6;1}^{\NeqFour}& (1^-,2^-, 3^+,4^-,5^+,6^+)\ 
=\
i \cg\ \left[ D_1\,\Wsix1 + D_2\,\Wsix2 + D_3\,\Wsix3 \right],
\cr}
\equn
$$
where
$$
\eqalign{
  D_1\ &=\ \left({ [3|K_{123}|4\ra
         \over t_{123} } \right)^4    \ B_0
       + \left({ \spa1.2\spa5.6 \over t_{123} } \right)^4
            \ \bar{B}_0 \ , \cr
  D_2\ &=\ \left({ [3|K_{234}|1\ra 
         \over t_{234} } \right)^4    \ B_0^+ 
       + \left({ \spa2.4\spb5.6 \over t_{234} } \right)^4
            \ \bar{B}_0^+   \ , \cr
  D_3\ &=\ \left(
{ [6|K_{345}|4\ra 
         \over t_{345} } 
\right)^4    \ B_0^- 
       + \left(
{ \spa1.2\spb3.5 \over t_{345} } \right)^4
             \bar{B}{}_0^-  , 
\cr}
\equn
$$
and
$$
\eqalign{
A_{6;1}^{N=4}& (1^-,2^+,3^-,4^+,5^-,6^+)
= i \cg\ \left[ G_1\,\Wsix1 + G_2\,\Wsix2 + G_3\,\Wsix3 \right],
\cr}
\equn
$$
where
$$
\eqalign{
  G_1\ &=\ \left({ [2|K_{456} | 5 \ra
         \over t_{123} } \right)^4    \ B_0
       + \left({ \spa1.3\spb4.6 \over t_{123} } \right)^4
            \ \bar{B}_0 \ , \cr
  G_2\ &=\ \left({ [6|K_{234} | 3 \ra 
 \over t_{234} } \right)^4    \bar{B}_0^+ 
       + \left({ \spa5.1\spb2.4 \over t_{234} } \right)^4
            \ B_0^+ \ , \cr
  G_3\ &=\ \left({ [4|K_{612} | 1\ra 
         \over t_{345} } \right)^4    \bar{B}_0^-
       + \left({ \spa3.5\spb6.2 \over t_{345} } \right)^4
            \ B_0^- \ . \cr}
\equn
$$

The seven-point ``split helicity'' NMHV amplitude, $A(1^-,2^-,3^-,4^+,5^+,6^+,7^+)$,  was computed in ref.~\cite{Britto:2004nj} 
and the remaining seven-point NMHV amplitudes in ref.~\cite{Bern:2004ky}, with the generalisation to all-$n$ being given in ref.~\cite{BDKn}.  The detailed formulae are given in these references, however the pattern for the NMHV is one of increasing complexity. For example the seven-point amplitude is
$$
A_7^{\text{NMHV}}= \sum_{i} c_i F^{1m}_{i} 
+\sum_{i} d_i F^{2me}_{3,i}
+\sum_{i} e_i F^{2mh}_{2,i}
+\sum_{i} f_i F^{2mh}_{3,i}
+\sum_{i} g_i F^{3m}_{2,2,i} \, ,
\equn
$$
which is a sum over all possible seven-point box functions. The formulae for the 35 coefficients are given in two different forms in ref.~\cite{Bern:2004ky} and ref.~\cite{BDKn}.  
Very few coefficients are absent. For example the amplitude $A(1^-,2^-,3^-,4^+,5^+,6^+,7^+)$ has 29 non-vanishing coefficients while in the amplitude
$A(1^-,2^+,3^-,4^+,5^-,6^+,7^+)$ all 35 coefficients are non-zero.  We have implemented the expressions of ref.~\cite{BDKn} and used them to study the seven- and eight-point NMHV amplitudes numerically.

\section{Numerical results for six-, seven- and eight-point amplitudes}

In this section we present a numerical study of Approximate Universality in $\NeqFour$ one-loop amplitudes.  Since four- and five-point amplitudes must be MHV (or conjugate to MHV) these amplitudes have an exact Universality; therefore  we start our study with the six-point amplitudes.  
The functional
forms of the different components of the six-point amplitudes are gathered together in~\cite{Dunbar:2009uk} together with the implementation of most of these in {\tt Mathematica}. 
These components  have also been calculated at a specific kinematic point and evaluated numerically by Ellis, Giele and Zandereghi (EGZ)~\cite{Ellis:2006ss}.  
For convenience we first examine the amplitudes at this kinematic point using the published numerical values. 
 
In table~\ref{sixpointN4} we extract the $\NeqFour$ amplitudes and split them into their $\eps^{-2}$, $\eps^{-1}$ and $\eps^0$ pieces.  
We also look at the combination 
$| { A^{\text{1-loop}} _{\eps^0} / A^{\rm tree} }|$.  If the amplitudes were exactly universal this combination would be identical for each helicity configuration. 
The numerical value of the tree amplitude is embedded within this table since the coefficient of $\eps^{-2}$ is just $6A^{\tree}$.  We have only
included a single MHV amplitude since the combination $| { A^{\text{1-loop}} _{\eps^0} / A^{\rm tree} }|$ is identical for the different MHV amplitudes.  
As we can see, despite the very different analytic forms of the NMHV expressions, the amplitudes are numerically almost universal.

\begin{table}[h]
\centering\begin{tabular}{ccccc}
\toprule 
  Amplitude  & $\eps^{-2}$        & $\eps^{-1}$   & 
$ \eps^{0} $ &  $| A^{\text{1-loop}}_{\eps^0}/A^{\rm tree} |$ \\
\midrule
 $\mmpppp $  &  $-161.917+54.826 i$      & $-489.024-212.415 i$
& $-435.281-1162.971 i$ & $43.584$ \\
$\mmmppp $  & $-6.478-10.4079 i$      &
 $6.825-37.620 i$ & $75.857-47.081 i$ & $43.698$ \\
$\mmpmpp $  & $14.074-22.9089 i$     &
 $80.503-23.464 i$ &  $169.047+93.601 i$ & $43.122$\\
$\mpmpmp $  &  $13.454+13.177 i$    &
 $13.454+13.177 i$ & $13.454+13.177 i$ & $43.311$ \\
 \bottomrule
\end{tabular}
\caption[]{The Almost Universal nature of the six-gluon $\NeqFour$ amplitudes at the EGZ kinematic point.}
\label{sixpointN4}
\end{table}

At this point we can also observe that Approximate Universality is only a property of the $\NeqFour$ amplitudes.  In table~\ref{thedarkside} we calculate
the combination $| { A^{\text{1-loop}} _{\eps^0} / A^{\rm tree} }|$ for $\NeqOne$ Yang--Mills and for  QCD/pure gauge.
Since we are looking at the leading in colour 
term, the massless quarks in the fundamental representation make no contribution so QCD is equivalent to the pure gauge theory. For $\NeqOne$ we examine
the vector multiplet without additional chiral multiplets. As we can see there is no strong universal behaviour. The amplitudes are roughly similar because the 
common gluonic contribution is the largest. 

\begin{table}[h]
\centering\begin{tabular}{ccc}
\toprule
  Amplitude  &  $\;\;\; |A^{\text{1-loop},\NeqOne}_{\eps^0}/A^{\rm tree}| \;\;\; $        &   $\;\;\; |A^{\text{1-loop},QCD}_{\eps^0}/A^{\rm tree}| \; $    \\
\midrule
 $\; A_6(--++++) \;$  &  43.8573    & 44.2379\\
 $\; A_6(-+-+++) \;$  &  45.5693    & 46.218 \\
 $\; A_6(-++-++) \;$  &   57.33    & 60.275 \\
$\; A_6(---+++) \;$  &   47.4327   &47.2043 \\
$\; A_6(--+-++) \;$  &    40.2536  &40.1239 \\
$\; A_6(-+-+-+) \;$  &     48.0032  &49.0462 \\
\bottomrule
\end{tabular}
\label{thedarkside}
\caption[]{The non-Universal nature of $\NeqOne$ and QCD  six-point amplitudes.}
\end{table}

\begin{figure}[h]
\centering
\includegraphics[totalheight=6.2cm]{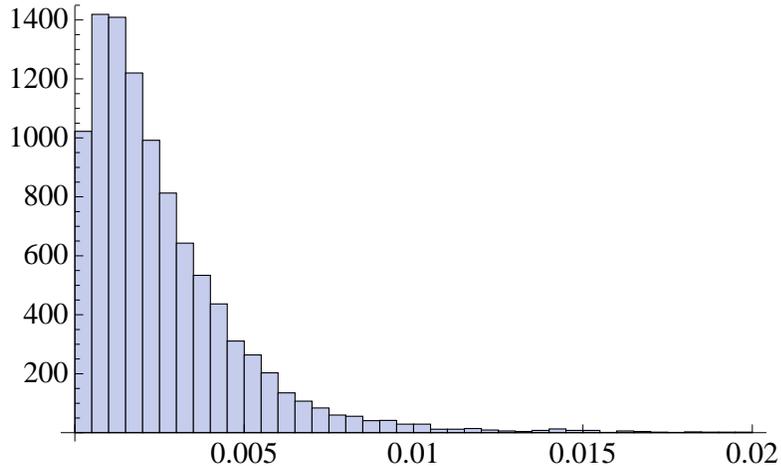}
\caption{Histogram of $10^4$ values of $\delta_H$ for amplitude $A_6(\mpmpmp)$.  }
\label{historesults}
\end{figure}

\begin{figure}[h!]
\centering
\includegraphics[totalheight=7.0cm]{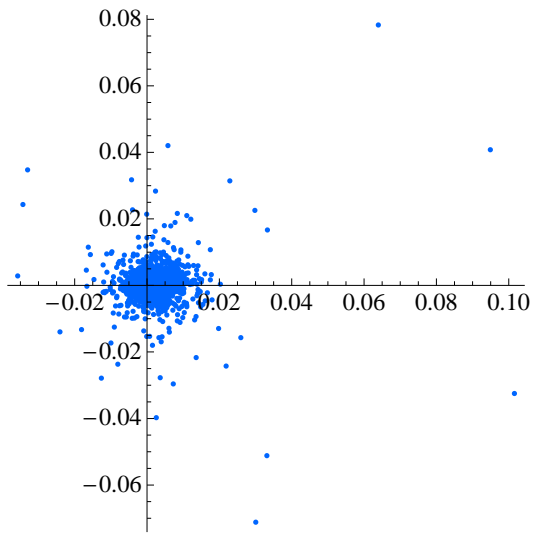}
\caption{${\cal E}_H/A^{\text{1-loop}}_{\eps^0}$ for $A_6(\mpmpmp)$ plotted in the complex plane for $10^4$ kinematic points.}
\label{complexplaneresults}
\end{figure}

The Approximate Universality of the six-point amplitudes evident at the EGZ kinematic point can be studied at other kinematic points and for higher point functions. 
We define the ``fractional error'' 
$$
\delta_H =  \left|{{ \cal E}_{H,n} \over  A^{\text{1-loop}}_{\eps^0}} \right|
= 
\left|{ A^{\text{1-loop}} -A^{\rm tree} U_n
\over  A^{\text{1-loop}}_{\eps^0} }\right|,
\equn
$$ 
as a measure of the deviation from exact universality, and in table~\ref{resultstable} we present the average value of $\delta_H$ for a set of $10^4$ random kinematic points for the six-, seven- and eight-point $\NeqFour$ NMHV amplitudes. As can be seen, Approximate Universality extends to random kinematic points and holds for seven- and eight-point NMHV configurations.  The average value $\bar\delta_H$ is a fraction of 1\% in all cases and is not obviously growing with the number of external legs. The spread of $\delta_H$ values for the amplitude $A_6(\mpmpmp)$ is illustrated in figure~\ref{historesults} and \ref{complexplaneresults};
for clarity in figure~\ref{historesults} we have excluded the very small number of outliers. For this amplitude, 1.8\% of kinematic points have $\delta_H > 0.01$,  0.05\% have $\delta_H >0.05$, and   0.03\% have $\delta_H >0.1$.

\begin{table}[h]
\centering\begin{tabular}{cc|cc|cc}
\toprule
 Amplitude  &  $ \bar\delta_H  $  &  Amplitude  &  $ \bar\delta_H  $ &
Amplitude  &  $ \bar\delta_H  $ \\
\midrule
$ A_6(\mmpppp) $  & 0  & $ A_7(\mmppppp) $  &   0  &    $ A_8 (\mmpppppp) $  &   0  \\
$ A_6(\mmmppp) $  &  0.0020  &$ A_7(\mmmpppp) $  &   0.0019   & $ A_8(\mmmppppp) $  &   0.0017   \\
$ A_6(\mmpmpp) $  &  0.0025 &  $ A_7(\mmpmppp) $  &   0.0030  & $ A_8(\mmpmpppp) $  &   0.0028  \\
$ A_6(\mpmpmp) $  & 0.0027  &$ A_7(\mmppmpp) $  &   0.0022   & $ A_8(\mmppmppp) $  &  0.0025   \\
$  $  &  &$ A_7(\mpmpmpp) $  &   0.0035   &  $ A_8(\mpmpmppp) $  &  0.0039    \\
      &  &   &     & $ A_8(\mpmppmpp)  $  &  0.0031   \\
\bottomrule
\end{tabular}
\caption[]{The Almost Universal nature of the six-, seven- and eight-gluon NMHV amplitudes. $\bar\delta_H$ is the average over $10^4$ kinematic points in each case.}
\label{resultstable}
\end{table}

In summary we see that the numerical evidence clearly shows the Approximately Universal nature of the $\NeqFour$ one-loop amplitudes.

\pagebreak

\section{$\NeqEight$ supergravity}

Maximal supergravity amplitudes share some features with those of maximal Yang--Mills theories.  
In particular, the one-loop amplitudes may also be expressed as a sum over scalar boxes~\cite{NoTriangle}.  There is, however,
 no concept of colour ordering and so there are many fewer helicity amplitudes for $n$-graviton scattering 
than there are for $n$-gluon scattering. 
For example at six and seven points there is just the MHV and a single  NMHV configuration.  This makes testing the concept of Universality rather limited. 
However, at six points we can  evaluate and compare $M^{\text{1-loop}}_{\eps^0}/M^{\text{1-loop}}_{\eps^{-1}}$ for both the MHV~\cite{Bern:1998sv} and NMHV\cite{BjerrumBohr:2005xx} amplitudes.  These are  proportional to $M^{\text{1-loop}}_{\eps^0}/M^{\rm tree}$ as the $\eps^{-1}$ term is  the tree up to an overall kinematic factor of 
$\sum s \ln(-s)$~\cite{DunNorB}. 
Results at the EGZ kinematic point are
shown in table~\ref{gravitytable}. Somewhat surprisingly the six-point amplitudes do not display an Approximately Universal behaviour. We have confirmed this at other kinematic points. 

\begin{table}[h]
\centering\begin{tabular}{ccccc}
\toprule
 Amplitude  &  $ \eps^{-1}$  &  $\eps^0$   & $M^{\text{1-loop}}_{\eps^0}/M^{\text{1-loop}}_{\eps^{-1}} $ \\     
\midrule 
 $ M_6(\mmpppp)$  &  $ 116\:110 + 1\:784\:376\,i$   & $ 974\:515 + 7\:216\:281\,i $    & $4.0624-0.2817\,i$ \\
$ M_6(\mmmppp) $  &  $1317.07 - 40.8019\,i $   & $ 4\:040.54 - 558.442\,i$ & $3.07802-0.32865\,i$ \\
\bottomrule
\end{tabular}
\caption[]{The six-point $\NeqEight$ supergravity amplitudes evaluated at the EGZ kinematic point do not show a Universal behaviour.}
\label{gravitytable}
\end{table}

\section{Conclusions}

There are many surprises but no accidents in Quantum Field Theory. We have been surprised by the almost universal nature of the one-loop amplitudes we have studied here. 
We have seen Approximate Universality in maximally supersymmetric gauge theory but not in pure gauge theory, minimally supersymmetric theories, nor $\NeqEight$ supergravity. 
The main contributions to the scattering appear completely determined by the tree amplitude. Explanations for this could be simply the amplitude being
largely determined by its infrared behaviour, or more intriguingly, an indication of a broken symmetry such as dual conformal symmetry~\cite{Drummond:2008vq}.   It would be interesting to explore this behaviour beyond one loop and possibly relate it to some of the other remarkable results in the $S$-Matrix of $\NeqFour$ super-Yang--Mills~\cite{Bern:2005iz,Brandhuber:2007yx}, and to explore whether 
Approximate Universality continues, or indeed becomes exact, in theories related to $\NeqFour$ Yang--Mills such as twistor string theory~\cite{Witten:2003nn}.

This research was supported by the STFC of the UK.

\end{document}